\documentstyle[12pt,epsfig] {article}
\begin{document}
\newcommand {\Zzero}     {\mathrm{Z}^0}
\newcommand {\MZ}        {M_{\mathrm{Z}}}
\newcommand {\GZ}        {\Gamma_{\mathrm{Z}}}
\newcommand {\MW}        {M_{\mathrm{W}}}
\newcommand {\GW}        {\Gamma_{\mathrm{W}}}
\newcommand {\GF}        {G_{\mathrm{F}}}
\newcommand {\Gm}        {G_{\mu}}
\newcommand {\thw}       {\theta_{\mathrm{W}}}
\newcommand {\swsq}      {\sin^2\!\thw}
\newcommand {\ee}        {\mathrm{e}^+\mathrm{e}^-}
\newcommand {\WW}      {\mathrm{W}^+\mathrm{W}^-}
\newcommand {\eeWW}    {\ee\to\WW}
\newcommand {\WWg}     {WW\gamma}
\newcommand {\WWZ}     {WWZ}
\newcommand {\eenunug} {\ee\to\nu\bar{\nu}\gamma}
\newcommand {\enuW}    {e^-\bar{\nu}_e \mathrm{W^+}}
\newcommand {\eetoenuW} {\ee\to\enuW}
\newcommand {\eetolnuW} {\ee\to\ell^-\bar{\nu}_{\ell}\mathrm{W}^+}
\newcommand {\eetoenumunu} {\ee\to e^-\bar{\nu}_e\mu^+\nu_{\mu}}
\newcommand {\thetaE}   {\theta_{e^-}}
\newcommand {\costhe}   {\cos\theta}
\def\thefootnote{\fnsymbol{footnote}}
\newif\ifPreprint\Preprinttrue

\ifPreprint
\begin{titlepage}
\small
\begin{flushright}
SAGA-HE-113 \\
October 1996\\
\end{flushright}
\large
\bigskip
\begin{center}
\Large\bf\boldmath 
Study on the single-W production \\
at LEP energies
\footnote{Invited talk at the Third International Symposium on
Radiative Corrections, Cracow, Poland, 1-5 August 1996.
To appear in Acta Physica Polonica.}
\end{center}
\begin{center}
\bigskip
\Large
Toshio Tsukamoto\\
\bigskip
\large
{\it Department of Physics, Saga University, Saga 840, Japan} \\
\end{center}                                                                    
\bigskip

\normalsize
\begin{center}{\large\bf Abstract}\end{center}
We present a study of single-W production ($\eetoenuW$)
as a new probe of the anomalous couplings at the LEP energy region.
The cross-section measurement of the single-W process 
is found to give complementary bounds on 
the anomalous couplings to those obtained from W-pair analysis
at LEP2.
\end{titlepage}
\fi
%
\bigskip
\begin{center}
\large\bf\boldmath 
Study on the single-W production \\
at LEP energies
\end{center}
\begin{center}
\large
Toshio Tsukamoto\\
\normalsize
{\it Department of Physics, Saga University, Saga 840, Japan} \\
\end{center}                                                                    

\small
\begin{abstract}
We present a study of single-W production ($\eetoenuW$)
as a new probe of the anomalous couplings at the LEP energy region.
The cross-section measurement of the single-W process 
is found to give complementary bounds on 
the anomalous couplings to those obtained from W-pair analysis
at LEP2.
\end{abstract}
\normalsize
\section{Introduction}
The non-abelian self couplings of gauge bosons are the most direct
consequences of the $SU(2)\times U(1)$ gauge symmetry and 
a direct confirmation of Triple Gauge boson Couplings (TGC)
is one of the most important subjects at LEP2.
Most studies up-to now have focused on the process $\eeWW$.
Although this process is anticipated to give a good sensitivity
to the anomalous couplings, it suffers from the disadvantage that 
one cannot disentangle the effects of $\WWg$ and $\WWZ$ couplings.
Especially, the gauge cancellations between $\gamma$, $\Zzero$ and $\nu$
exchange graphs are still not fully operative at LEP2, hence, 
only the interference effects between different TGCs dominate.

One way to avoid such complications is to use the channels which is
only sensitive to the $\WWg$ coupling.
Recently we have proposed a new probe of the $\WWg$ vertex using
$\eetoenuW$ process where the electron escapes down the beam pipe
~\cite{fullpap}.
In this report, we give a brief summary of the characteristics of the
single-W production and present the anticipated sensitivity to 
the anomalous TGCs.
\section{Single-W production and its signature}
\begin{figure}[thb]
\vspace*{-1cm}
\begin{center}\hspace*{-.5cm}
\mbox{
\epsfxsize=12cm\epsfysize=8.cm\epsffile{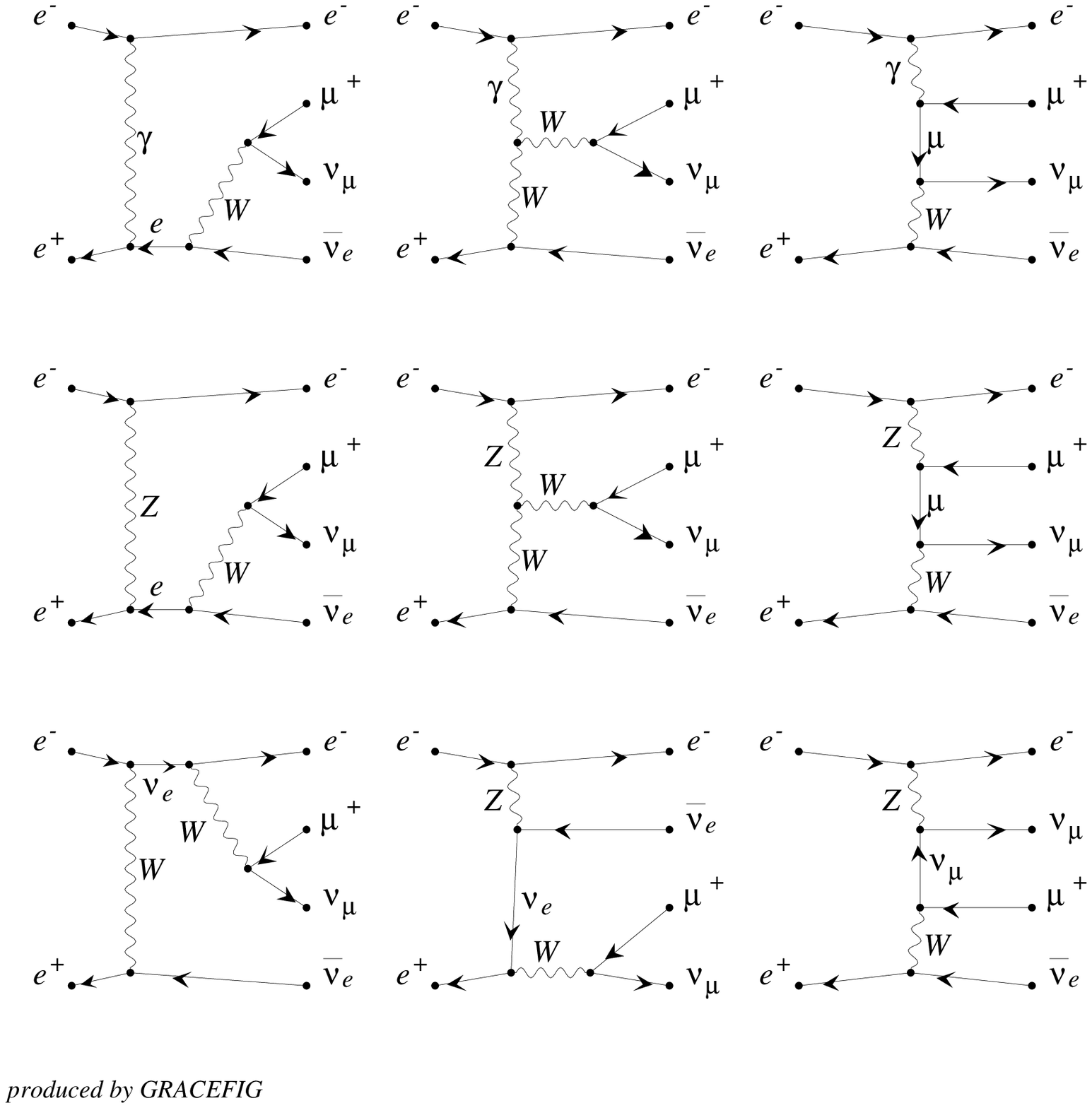}}
\caption{\small\sf
The $t$-channel diagrams of the $\eetoenumunu$ process.}
\label{f-diag2}
\end{center}
\vspace*{-.7cm}
\end{figure}

Figure~\ref{f-diag2} shows a gauge invariant set of $t$-channel diagrams
involved in $\eetoenumunu$.
Among them, the $\gamma$-$W$ process (the first row)
give the dominant contribution.
Even below the WW threshold, a significant contribution is expected
from the $\gamma$-$W$-$W$ diagram (the second graph).

The contribution of the $t$-channel can be enhanced by requiring
the outgoing electron to be within a small angle. 
To calculate the cross-section with this requirement,
special care has to be taken of the following points;
(1) the matrix element needs to keep even the electron mass
finite throughout the calculation to avoid a singularity at small
electron polar angle ($\thetaE$). 
(2) the gauge violating term due to the introduction of the finite width
of W is found to blow up at small $\thetaE$~\cite{Gviol-1},
and stops one from obtaining
reliable cross-section as a result. This problem can, however, be
avoided by several methods~\cite{Gviol-2,Gviol-3} 
and the results of each scheme are found to be consistent~\cite{Gviol-3}. 

Here we use a four-fermion generator ``grc4f''~\cite{grc4f} for the calculation. 
In this package,
all the fermion masses are properly taken into account
and the gauge violation due to the finite width of
the W is cured~\cite{Gviol-2}.
One can thus perform a cross-section calculation reliably even without
a cut on the electron polar angle.

\begin{figure}[htb]
\vspace*{-.5cm}
\begin{center}\hspace*{-.5cm}
\mbox{
\epsfxsize=12.5cm\epsfysize=6.cm\epsffile{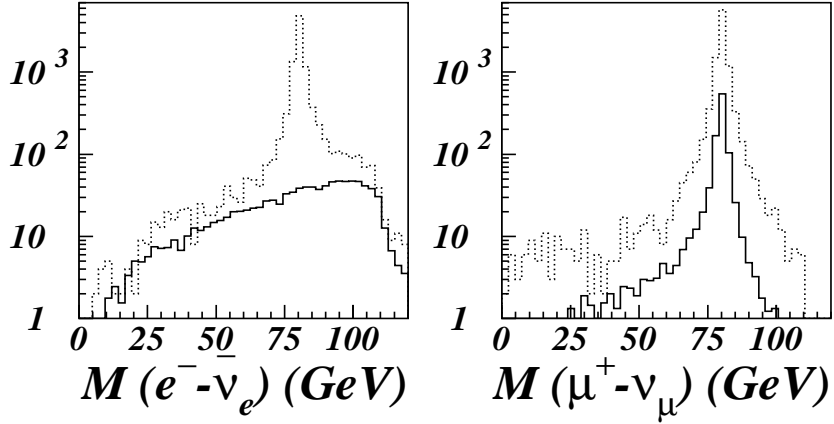}}
\caption{\small \sf 
$M(e^-$-$\bar{\nu_{e}})$ and $M(\mu^+$-$\nu_{\mu})$  distributions 
for $\eetoenumunu$ process
with no cut (dotted line) and single-W cuts (solid line).}
\label{f-evdist}
\end{center}
\vspace*{-.7cm}
\end{figure}

We mainly focus on the $\eetoenumunu$ process in this report 
since the signature is very clean from an experimental point of view.
To discriminate single-W from WW, we introduce ``single-W cuts'':
$\thetaE <$ 35 mrad, $| \costhe_{\mu^+} | <$ 0.95 and $P_{\mu} >$ 20 GeV.
The resulting event signature is high P$_t$ ``single muon''.
As is seen in figure~\ref{f-evdist}, double resonant
contribution is clearly suppressed while keeping single-W.

The total cross-section of the single-W process 
with $e\nu\mu\nu$ final state is shown in table~\ref{tab:xsect}.
One can observe ``single muon'' events from single-W production
with modest luminosity at LEP2.

\begin{table}[hbt]
\vspace*{-0.5cm}
\begin{center}
\begin{tabular} {|c|c|c|c|c|}
\hline
$E_{cm}$ (GeV) & 161 & 176 & 188 &  192 \\
\hline
$\sigma$ (fb)  & 28  & 39  & 50  &   54 \\
\hline
\end{tabular}     
\caption[foo]
{\label{tab:xsect}
Cross-section of single-W process with $e\nu\mu\nu$ final state }
\end{center}  
\vspace*{-1.2cm}
\end{table}

\section{Anomalous couplings}
To test the triple gauge boson coupling, we use the following effective
Lagrangian~\cite{TGCeffLag} assuming both $C$ and $P$ conservation:
\begin{eqnarray}
 i {\cal L}_{eff}^{WWV} = &g_{WWV} \Bigl[ \,
    g_1^V \left( W^{\dagger}_{\mu\nu}W^{\mu}V^{\nu} 
               - W^{\dagger}_{\mu}V_{\nu}W^{\mu\nu} \right)
    \nonumber \\
 &+ \kappa_V W^{\dagger}_{\mu} W_{\nu} V^{\mu\nu} 
 + \frac{\lambda_V}{m^2_W}  W^{\dagger}_{\rho\nu} W^{\mu}_{\nu} V^{\rho\nu} 
     \,\Bigl]
\end{eqnarray}
\noindent
where $V = \gamma$ or Z, and the overall couplings are
$g_{WW\gamma} = e$, $g_{WWZ} = e \cot \theta_W$, 
$W_{\mu\nu} = \partial_{\mu} W_{\nu} - \partial_{\nu} W_{\mu}$
and $ V_{\mu\nu} = \partial_{\mu} V_{\nu} - \partial_{\nu} V_{\mu}$.

Since $g_1^{\gamma}$ is required to be 1 by electromagnetic gauge
invariance, deviations from the Standard Model are defined as 5 parameters:
\begin{eqnarray}
\vspace*{-0.4cm}
   \Delta g^Z_1 \equiv (g^Z_1-1), \>\>
   \Delta \kappa_{\gamma} \equiv (\kappa_{\gamma} -1), \>\>
   \Delta \kappa_{Z} \equiv (\kappa_{Z} -1), \>\>
   \lambda_{\gamma}, \>\> \lambda_Z
   \label{eq:tgc1}
\vspace*{-0.4cm}
\end{eqnarray}

The anomalous TGCs are already severely constrained by low energy 
data~\cite{TGCthLOW}.
The parameters in equation (\ref{eq:tgc1}) are no longer independent
each other in order to protect low energy observables 
from acquiring discrepancies with the experimental data~\cite{TGCthSUM1}.
In case the $SU(2)_L\times U(1)_Y$ gauge symmetry is realized linearly,
only three of the five couplings are found
to be independent~\cite{TGCthLOW,TGCthSUM1}.
As a result,
the $WWZ$ couplings are related to the $WW\gamma$ ones with the equations:
\begin{eqnarray}
\vspace*{-0.4cm}
\Delta\kappa_{\gamma} = -\cot^2\theta_W\cdot(\Delta\kappa_Z - \Delta g^Z_1),
 \quad\quad \lambda_{\gamma} = \lambda_Z.
 \label{eq:tgc2}
\vspace*{-0.4cm}
\end{eqnarray}
The anticipated best sensitivity from the W-pair analysis at LEP2 are 
inferred based on this relation, which are found to be~\cite{TGC-genr}: 
$\Delta\kappa_{\gamma} = 0.06$, $\lambda_{\gamma} = 0.04$ and 
$\Delta g^Z_1 = 0.02$.

\begin{figure}[htb]
\vspace*{-.5cm}
\begin{center}\hspace*{-1.0cm}
\mbox{
\epsfxsize=12.cm\epsfysize=8.cm\epsffile{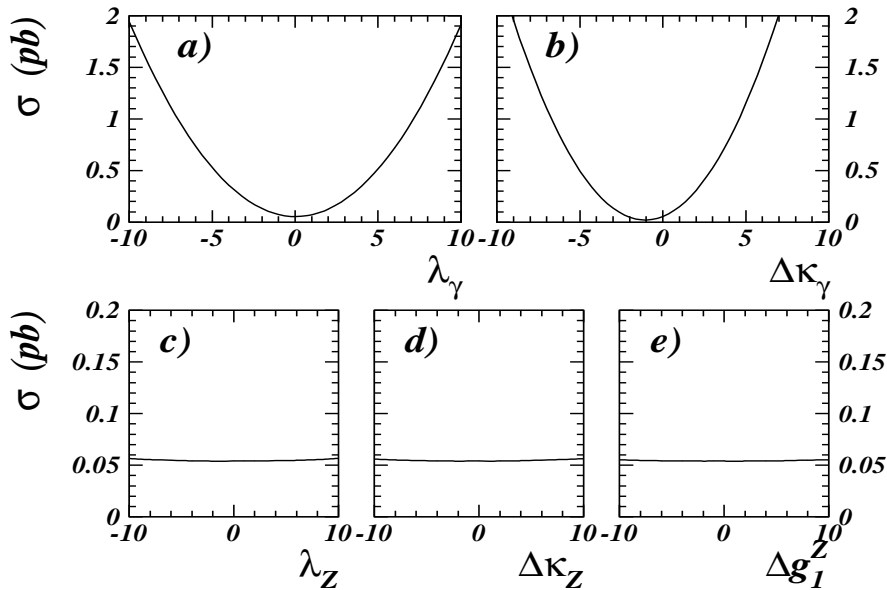}}
\caption{\small\sf
Variation of the cross-section for the single-W process 
($e\nu\mu\nu$).
Note that the vertical scales are expanded for Z-related couplings (c-e).
}
\label{f-ACgamz}
\end{center}
\vspace*{-.7cm}
\end{figure}

In contrast to the W-pair production, the observable of the single-W
process is less sensitive to these relations
since the contribution from the $WWZ$ vertex diagram for single-W process
is very small at LEP energies.
We demonstrate the variation of the cross-section at $E_{cm}=$ 192~GeV
as a function of each anomalous coupling in figure~\ref{f-ACgamz}.
The cross-section depends only marginally on $WWZ$ related couplings
while we see a large sensitivity to $WW\gamma$ ones.
This orthogonal feature against $WWZ$ couplings not only helps to extract 
anomalous $WW\gamma$ couplings alone but also constrain $WWZ$ couplings
once the W-pair analysis restricts the parameter space along equation 
(\ref{eq:tgc2}).
\begin{figure}[htb]
\vspace*{-.5cm}
\begin{center}\hspace*{-.5cm}
\mbox{
\epsfxsize=12cm\epsfysize=8.cm\epsffile{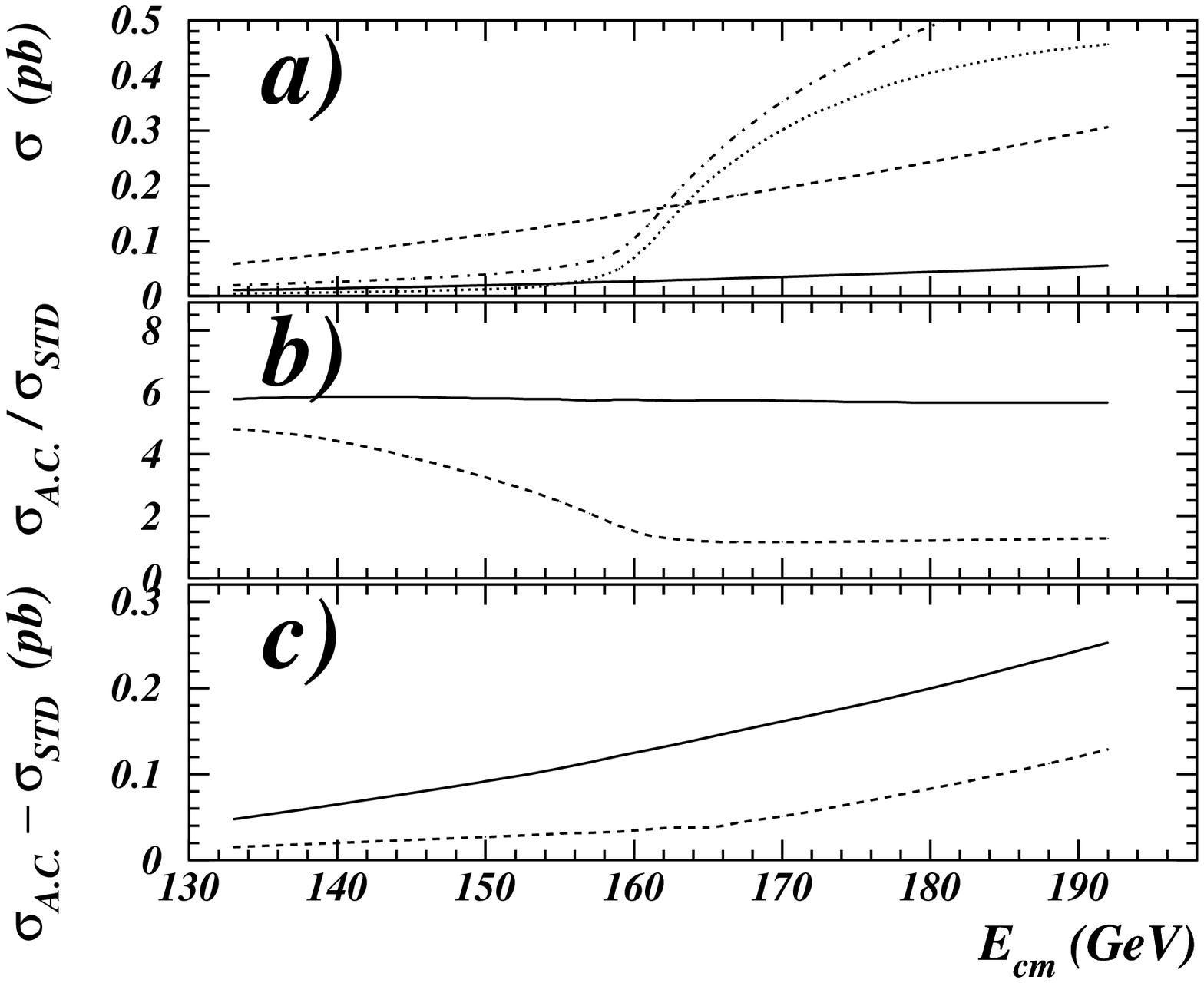}}
\caption{\small \sf
The enhancement of the cross-section due to the anomalous couplings
as a function of $E_{cm}$.}
\label{f-AC1w_ww}
\end{center}
\vspace*{-.7cm}
\end{figure}

Figure~\ref{f-AC1w_ww}-a), b) and c) illustrate a specific 
sensitivity to the anomalous coupling as a function of $E_{cm}$.
In figure~\ref{f-AC1w_ww}-a), the solid line corresponds to
the Standard Model cross-section with ``single-W'' cut.
If we assume anomalous couplings to be 
$\Delta \kappa_{\gamma}  =-\cot^2\theta_W\cdot\Delta\kappa_Z = 2$, 
we get a significantly larger cross-section (dashed line).
Also in the figure, the ratio of two cases (b) and
the difference between two (c) are shown as a function of $E_{cm}$.
The enhancement factor of about 6 is expected,
which is in marked contrast to W-pair production since
the measurement of the WW total cross-section is not sensitive
to the anomalous coupling around the LEP energy region.
We give the sensitivity 
for W-pair production as dashed lines in figure~\ref{f-AC1w_ww}-b) and c), 
where the ``canonical cuts''~\cite{MC4fermi} are made 
to enhance WW contribution.
The W-pair case is much less sensitive than the single-W case.

The anticipated sensitivity based on 1-parameter fit 
for the anomalous couplings are;
$ -0.4 < \Delta \kappa_{\gamma} < 0.3$ and $-0.9 < \lambda_{\gamma} < 1.0$,
at 95\% C.L., where $E_{cm}=$ 192~GeV and $\int{\cal L}dt =$ 500pb$^{-1}$ are
assumed. We emphasise that one can improve the limits further by including
other leptonic and hadronic channels.
The hadronic channel is especially very attractive
because of its large cross-section ($\sim$350fb at $E_{cm}=$ 192~GeV).
Single-W process is, in general, more sensitive to $\Delta\kappa_{\gamma}$
than to $\lambda_{\gamma}$. Although the anticipated bound on  
$\lambda_{\gamma}$ will not be attractive ($\sim 0.6$), 
the sensitivity of $|\Delta\kappa_{\gamma}| \sim 0.1$ is expected 
with $\int{\cal L}dt =$ 500pb$^{-1}$ which is comparable to 
that will be obtained from W-pair studies~\cite{TGC-genr}.
\section{Conclusion}
We have for the first time presented the TGC studies
making use of single-W production at the LEP energy region. 
The cross-section measurement of this process 
are found to give good sensitivities to the anomalous couplings,
in particular to $\Delta\kappa_{\gamma}$.

We emphasise that the precise study of the $WW\gamma$ vertex 
from $\eetoenuW$ process is important to disentangle the complex effects
coming from $WW\gamma$ and $WWZ$ vertices
which will be obtained from W-pair production analyses.
In this sense, the bounds from single-W process are complementary
to W-pair ones.

\vspace{0.5cm}
\noindent
{\bf Acknowledgements:}
The author would like to thank 
Minami-Tateya theory group of KEK for their 
physics discussions and technical help in grc4f system. 
This work was supported in part by Japan Society for Promotion of Science
under Fellowship for Research Program at Centres of Excellence Abroad.
%

%

%
\end{document}